\documentclass[a4paper,
              ]{jacow}
\makeatletter%
	\ifboolexpr{bool{xetex}}
	 {\renewcommand{\Gin@extensions}{.pdf,%
	                    .png,.jpg,.bmp,.pict,.tif,.psd,.mac,.sga,.tga,.gif,%
	                    .eps,.ps,%
	                    }}{}
\makeatother

\ifboolexpr{bool{xetex} or bool{luatex}} 
 {}                                      
 {\usepackage[utf8]{inputenc}}           
 \usepackage[T1]{fontenc}   
\usepackage[british,english]{babel}
\usepackage[final]{pdfpages}
\usepackage{multirow}
\usepackage{ragged2e}
\usepackage[caption=false]{subfig}
\usepackage{graphicx,dblfloatfix}
\graphicspath{{figures/}}

\ifboolexpr{bool{jacowbiblatex}}%
 {%
  \addbibresource{jacow-test.bib}
  \addbibresource{biblatex-examples.bib}
 }{}
\begin{document}

\title{ASSESSMENT OF TRANSVERSE INSTABILITIES IN PROTON DRIVEN HOLLOW PLASMA WAKEFIELD ACCELERATION\thanks{This work is supported by the PDS Award of The University of Manchester, the Cockcroft Institute core grant, and STFC. The authors greatly appreciate the computing resources provided by STFC Scientific Computing Department's SCARF cluster.}}

\author{Y. Li\thanks{yangmei.li@manchester.ac.uk}\textsuperscript{1,2}, G. Xia\textsuperscript{1,2}, Y. Zhao\textsuperscript{1} \\ 
\textsuperscript{1}University of Manchester, Manchester, UK \\ 
\textsuperscript{2}Cockcroft Institute, Daresbury, UK}
	
\maketitle

\begin{abstract}
Hollow plasma has been introduced into the proton-driven plasma wakefield accelerators to overcome the issue of beam quality degradation caused by the nonlinear transverse wakefields varying in radius and time in uniform plasma. It has been demonstrated in simulations that the electrons can be accelerated to energy frontier with well-preserved beam quality in a long hollow plasma channel. However, this scheme imposes tight requirements on the beam-channel alignment. Otherwise asymmetric transverse wakefields along the axis are induced, which could distort the driving bunch and deteriorate the witness beam quality. In this paper, by means of the 2D cartesian particle-in-cell simulations, we examine the potentially detrimental effects induced by the driving beam-channel offset and initial driver tilt, and then propose and assess the solutions to these driver inaccuracy issues.  
\end{abstract}

\section{INTRODUCTION}

Proton driven plasma wakefield acceleration (PD-PWFA) has attracted numerous attention since its inception, as the proton drivers with huge energies have the potential of driving wakefields for hundreds or even thousands of meters and accelerating the witness bunch to TeV energies \cite{item:1,item:2,item:3}. However, unlike electron drivers or lasers, the protons attract plasma electrons from different radii towards the axis instead of expelling them simultaneously. Therefore, the plasma electrons cross the axis at different times, leading to different phases. In this case, there are still some plasma electrons distributing in the accelerating bubble. The resulting transverse fields are radially nonlinear and also vary along the witness bunch and with the propagation distance, which will degrade the witness beam quality.

In our previous work, hollow plasma channels have been employed to resolve this issue \cite{item:4,item:5}. The hollow channel enables the witness electron bunch to reside in a region, which is completely free from plasma electrons and ions. Therefore, no transverse plasma wakefields act on the witness bunch. It is only focused by the external weak quadrupole field and the beam quality is well preserved. The issue in the hollow scheme is it requires perfect alignment of the driver, the witness bunch and the hollow channel on axis. But in reality, with laser jitters or other unexpected factors, there might be beam-channel misalignment or beam tilt. In this paper, we examine the transverse instabilities induced by the imperfectly aligned driver and demonstrate how the driver and the witness bunch evolve. We also propose solutions to mitigate the instabilities.

\section{SIMULATIONS}
  
\begin{table}[!b]
 \vspace*{-.8\baselineskip}
   \centering
   \caption{Proton and Plasma Parameters}
   \begin{tabular}{lcc}
       \toprule
       \textbf{Parameters} &\textbf{Values}                      &\textbf{Units} \\
       \midrule
           Bunch population, $N_p$                         & \SI{1.15}~$\times$~$10^{11}$              &               \\ 
           Bunch energy, $W_0$                        & \SI{1}                                            &TeV        \\ 
           Energy spread, $\delta W/W$              & 10$\%$                                           &                  \\ 
           RMS length, $\sigma_x$                 & \SI{150}                                                &$\mu$m              \\
           RMS radius, $\sigma_y$                 & \SI{350}                                             &$\mu$m      \\
           Normalized emittance, $\epsilon_n$              & \SI{11}                                               &$\mu$m                 \\
           Plasma density, $n_e$                         & \SI{5}~$\times$~$10^{14}$                &cm$^{-3}$  \\
           Hollow channel radius, $y_c$              & \SI{350}                                             &$\mu$m      \\
       \bottomrule
   \end{tabular}
   \label{table1}
  \vspace*{-1\baselineskip}
\end{table}

\begin{figure*}[!t]
  \vspace*{-1.8\baselineskip}
 \begin{center}
  \subfloat{
   \centering
   \includegraphics[width=0.25\linewidth]{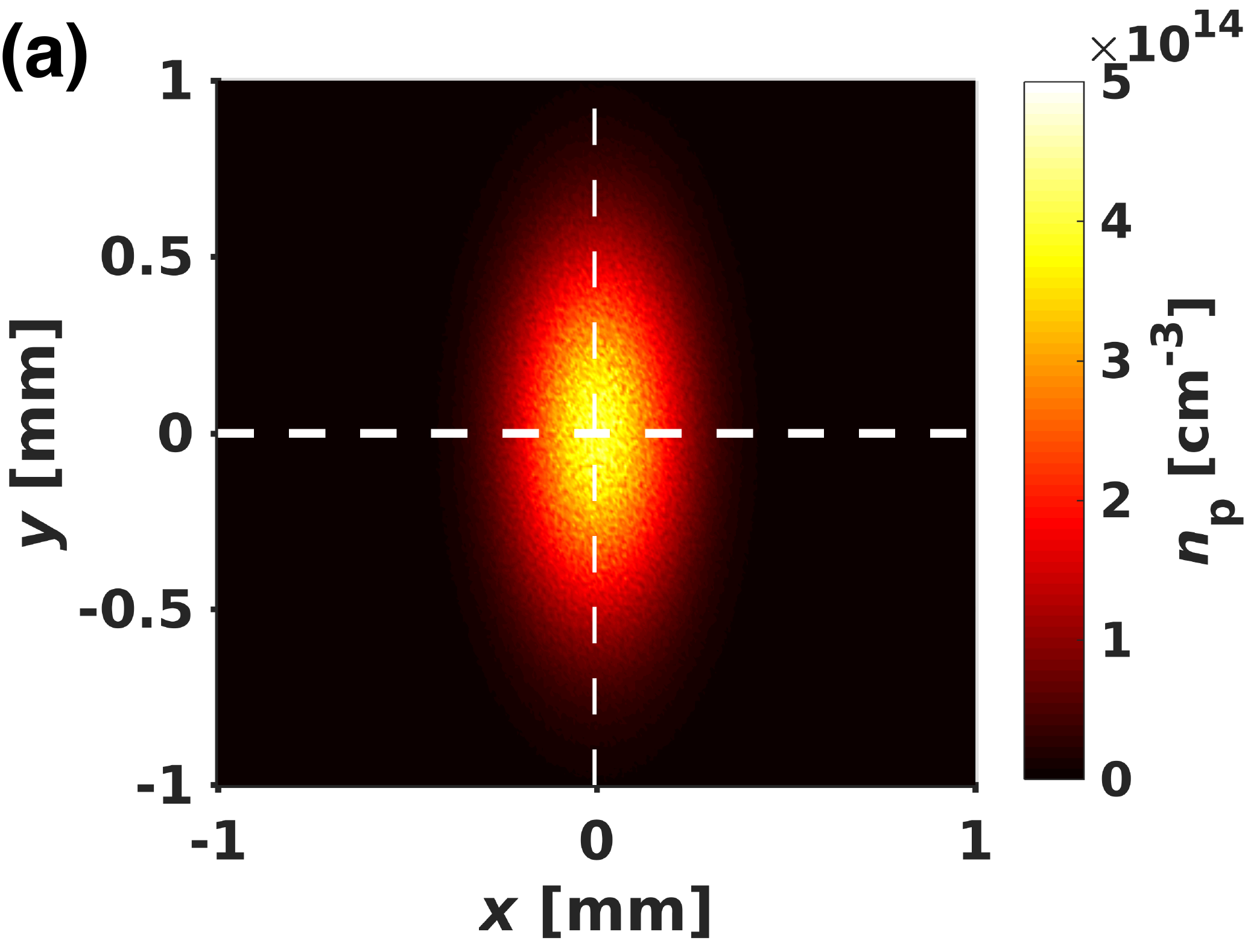}
  }
  \subfloat{
   \centering
   \includegraphics[height=3.25cm, width=0.375\linewidth]{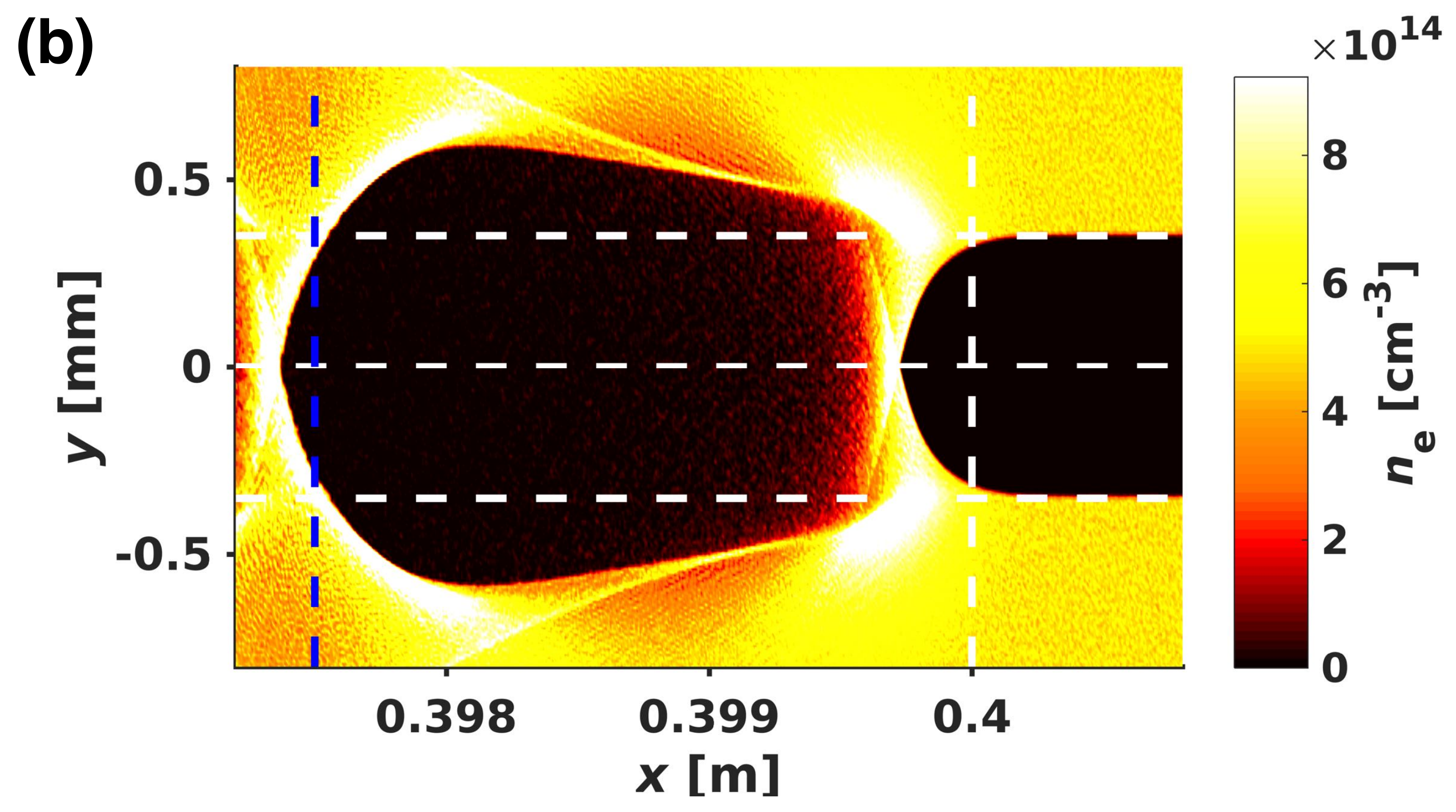}
  }
    \subfloat{
   \centering
   \includegraphics[height=3.25cm,width=0.375\linewidth]{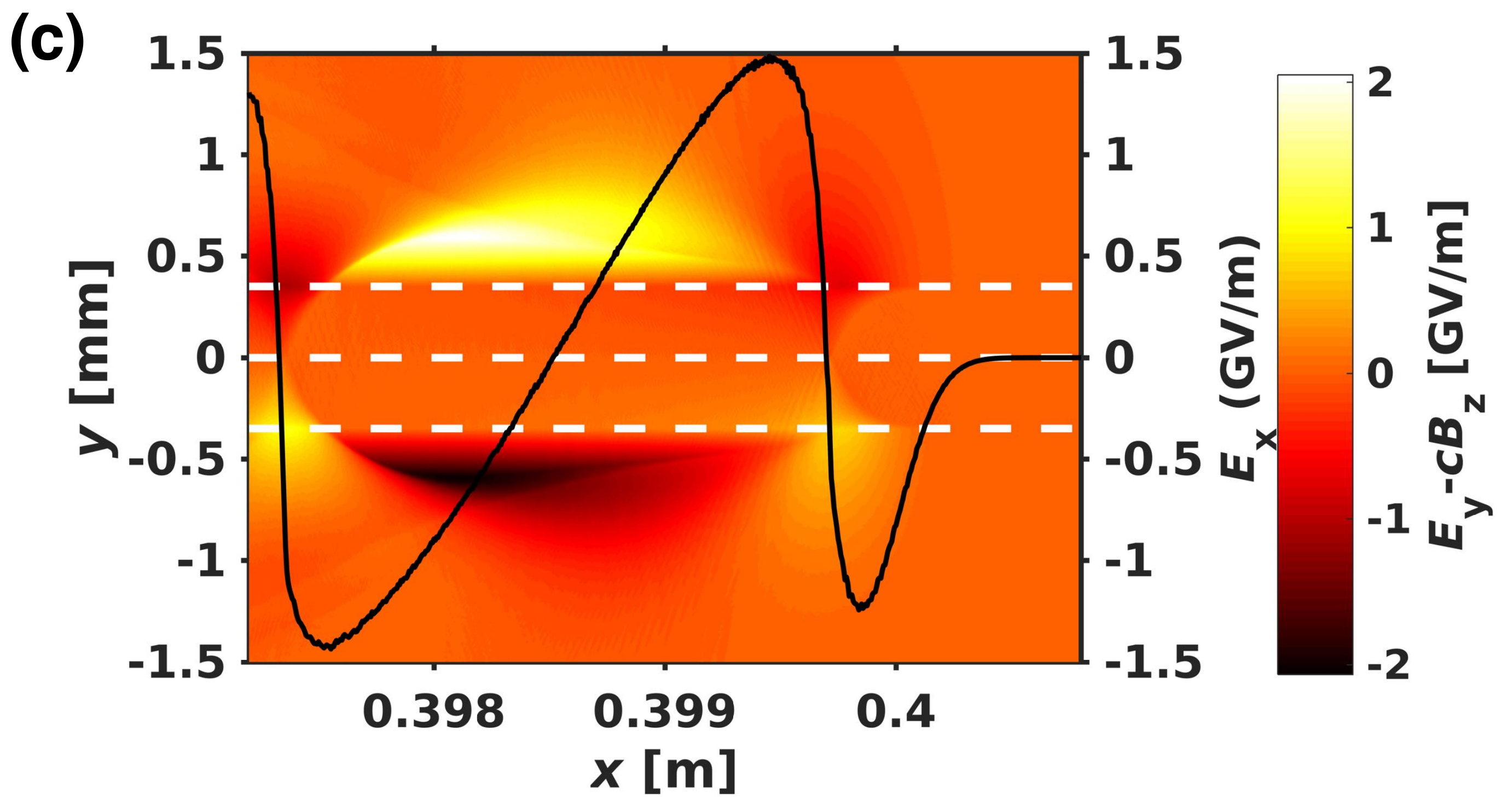}
  }
  \newline
  \subfloat{
   \centering
   \includegraphics[width=0.25\linewidth]{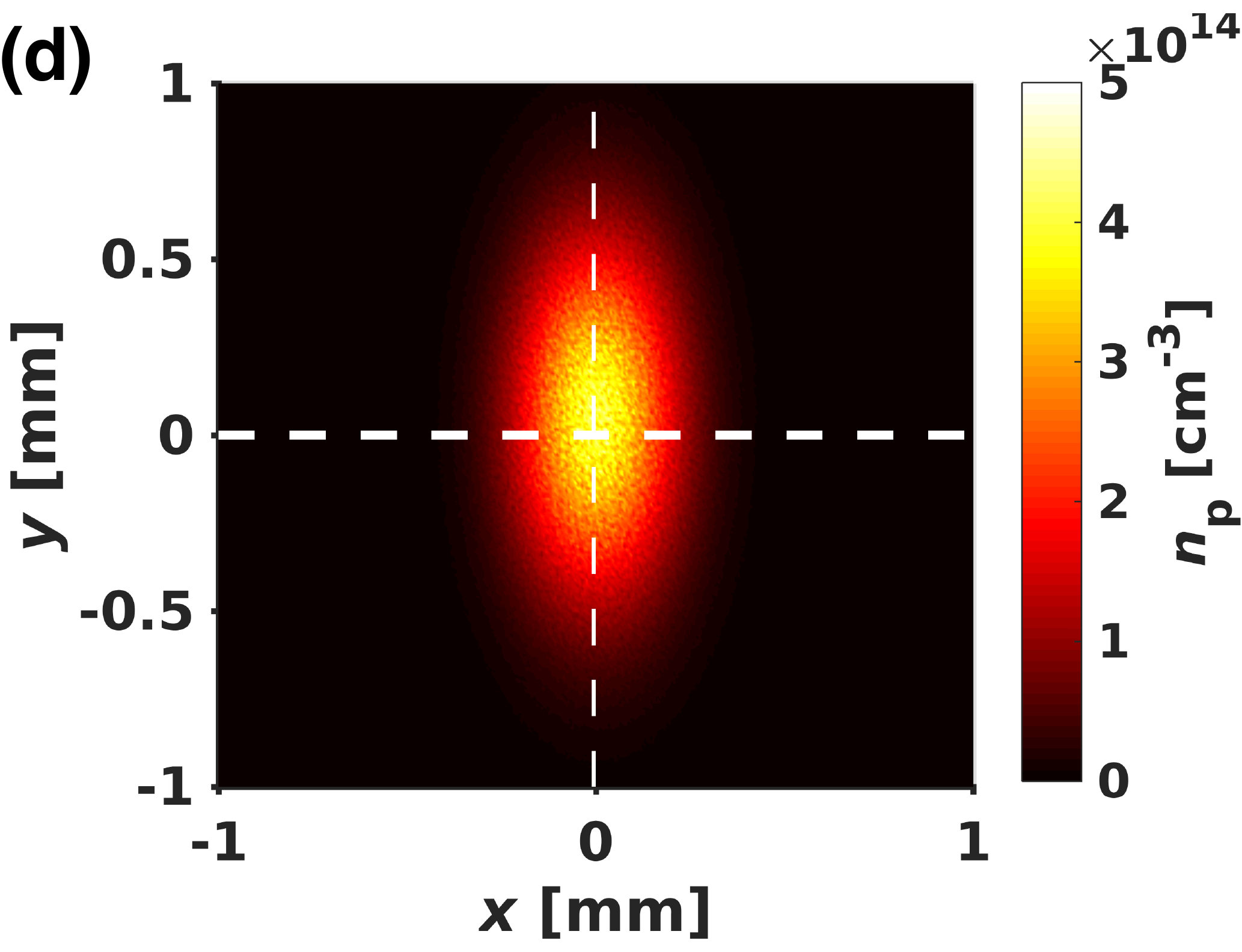}
  }
  \subfloat{
   \centering
   \includegraphics[height=3.25cm,width=0.375\linewidth]{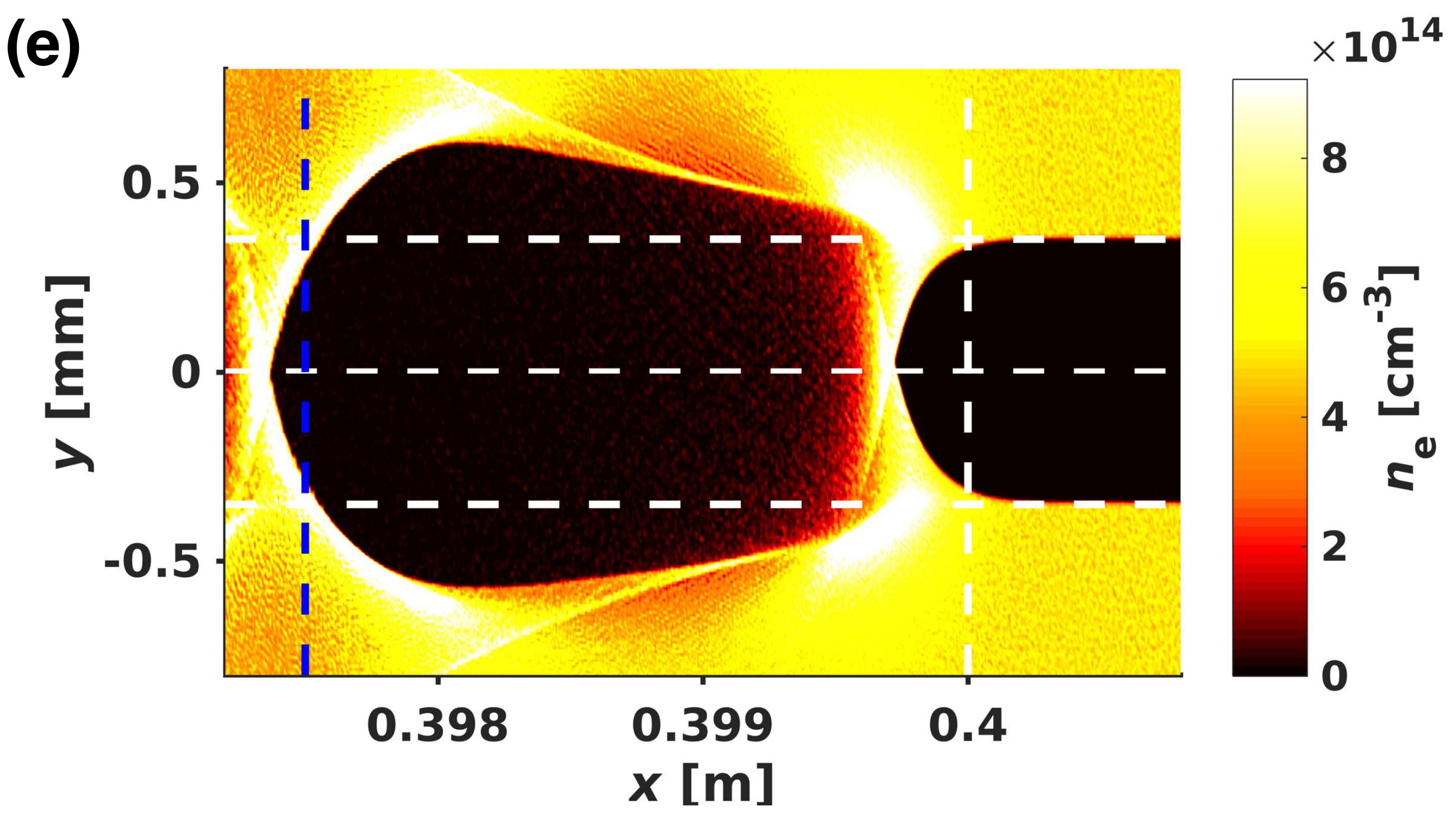}
  }
    \subfloat{
   \centering
   \includegraphics[height=3.25cm,width=0.375\linewidth]{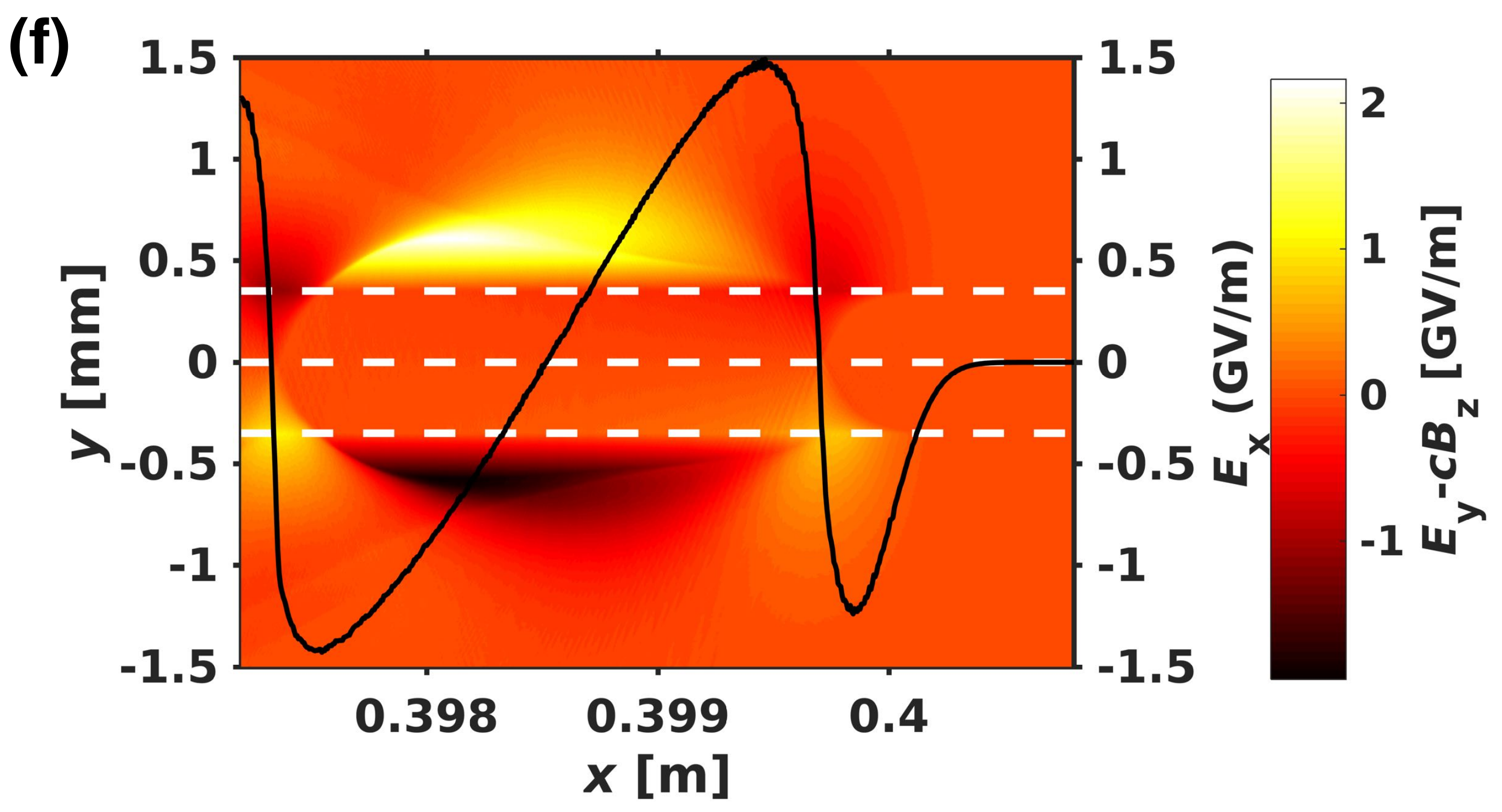}
  }
 \newline
  \subfloat{
   \centering
   \includegraphics[width=0.25\linewidth]{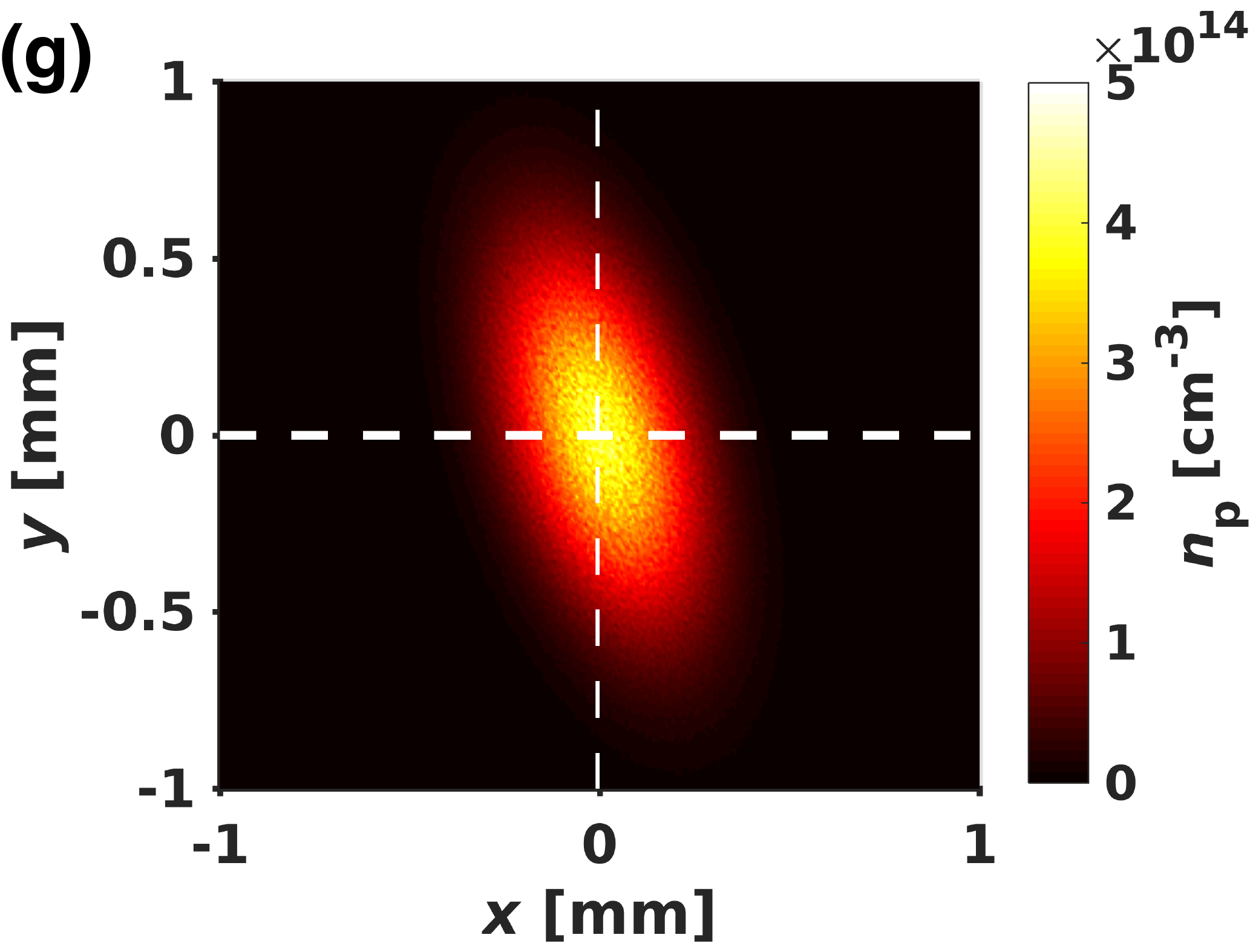}
  }
  \subfloat{
   \centering
   \includegraphics[height=3.25cm,width=0.375\linewidth]{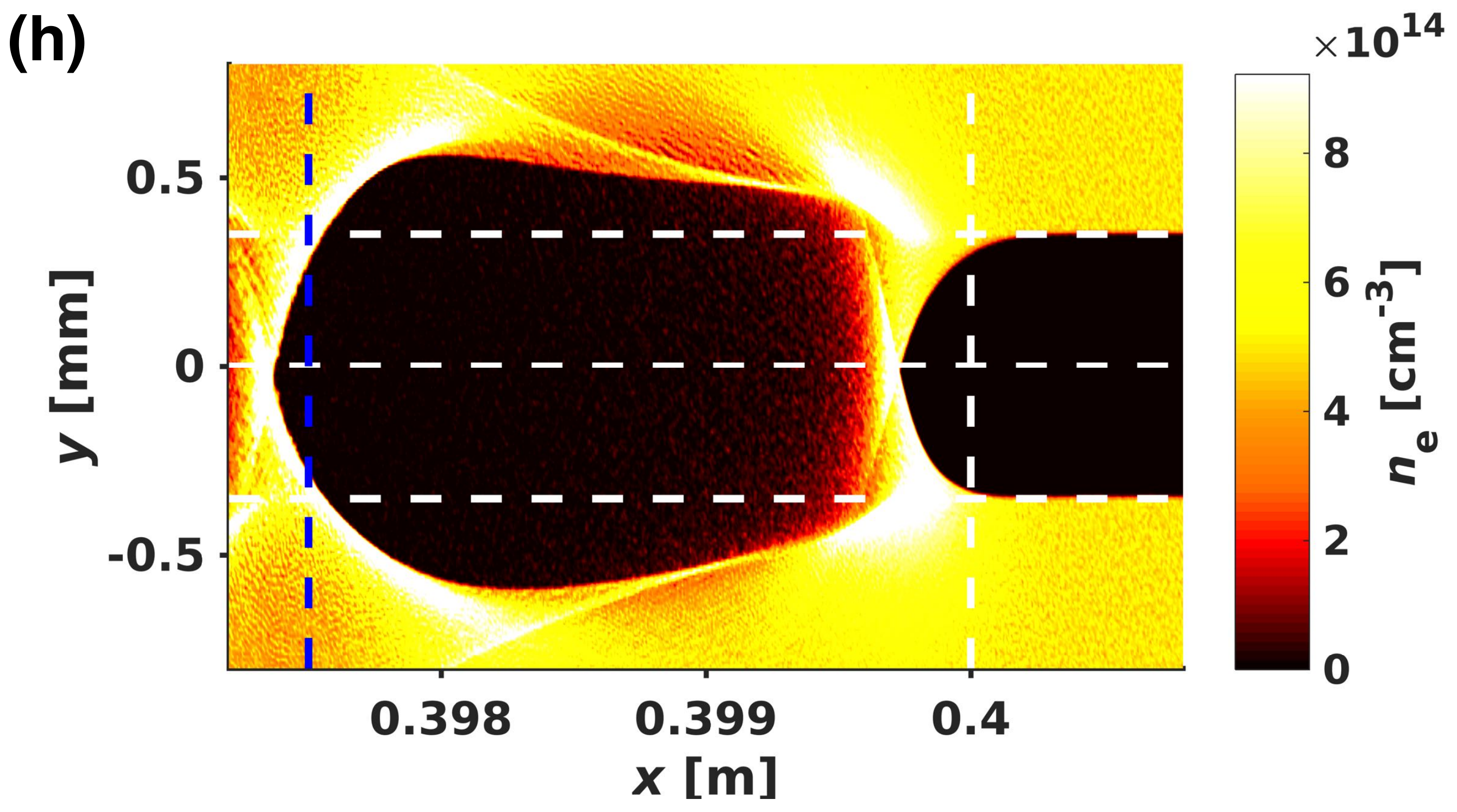}
  }
    \subfloat{
   \centering
   \includegraphics[height=3.25cm,width=0.375\linewidth]{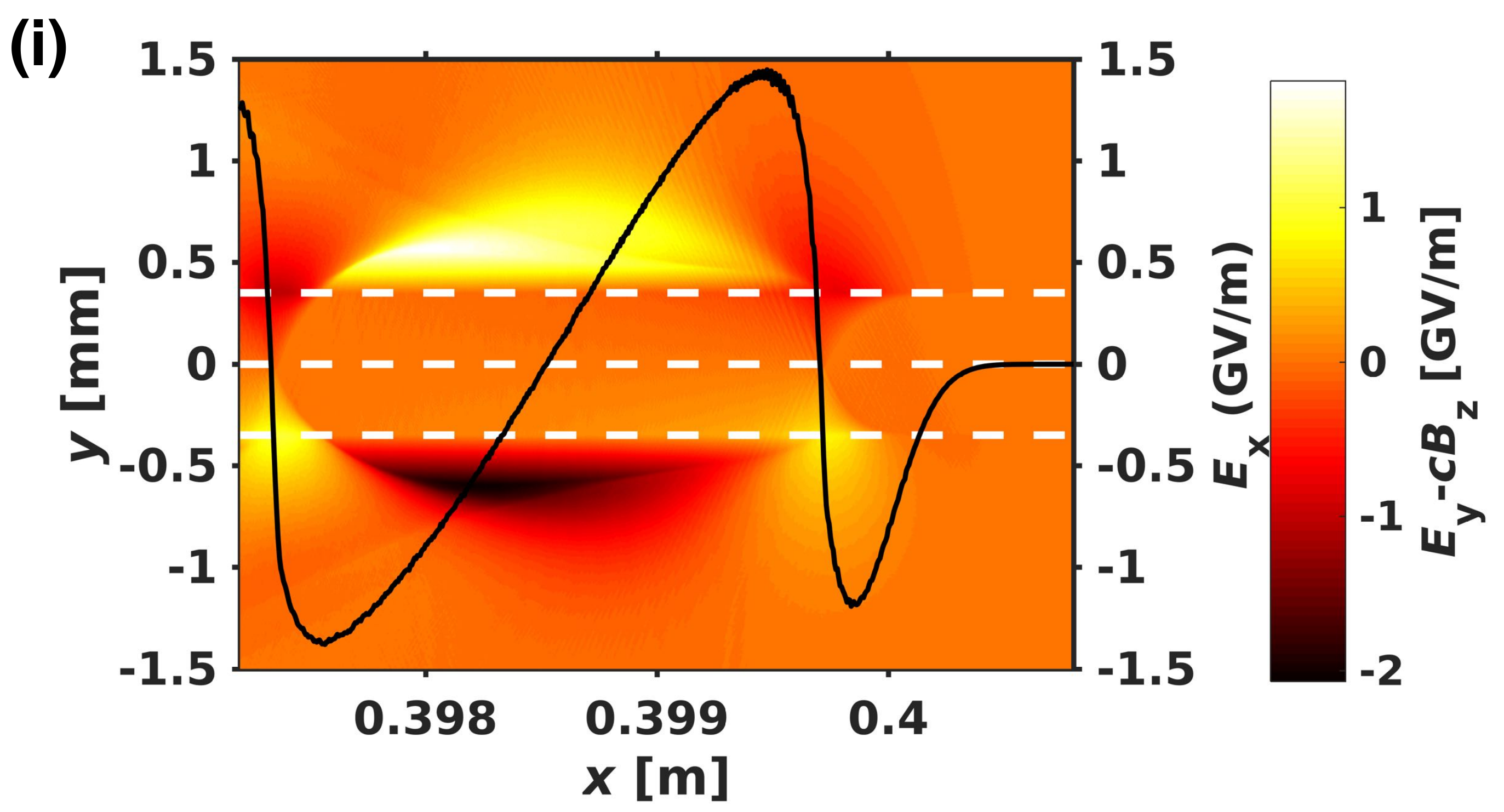}
  }
 
 \end{center}
 \caption{Initial proton driver profiles (1st column), plasma electron density distribution (2nd column) and transverse plasma wakefield distribution (3rd column) along with the on-axis longitudinal electric field (black curve) when the proton bunch travels for 0.4 m for three considered cases. The white and blue vertical dashed lines indicate the longitudinal centre of the driver and the the witness bunch, respectively. The white horizontal dashed line along $y=0$ marks the longitudinal axis while the other two off-axis ones mark the hollow channel boundaries.}
 \label{comparison}
  \vspace*{-\baselineskip}
\end{figure*}

With the open source particle-in-cell code EPOCH \cite{item:6} based on a 2D cartesian geometry, we first study the beam dynamics of the proton driver initially off-axis or tilted and then demonstrate how the driver misalignment or tilt affects the wakefields in the witness area. Afterwards, we propose three potential solutions especially the one employing a near-hollow plasma structure to confine the diffracted witness bunch. The simulation parameters for protons and plasma are given in Table~\ref{table1}. Note that we only consider the case with a single and short proton driver here. 

\subsection{Beam Dynamics of the Proton Driver}
 \begin{figure}[!h]
 \begin{center}
  \subfloat{
   \centering
   \includegraphics[width=0.47\linewidth]{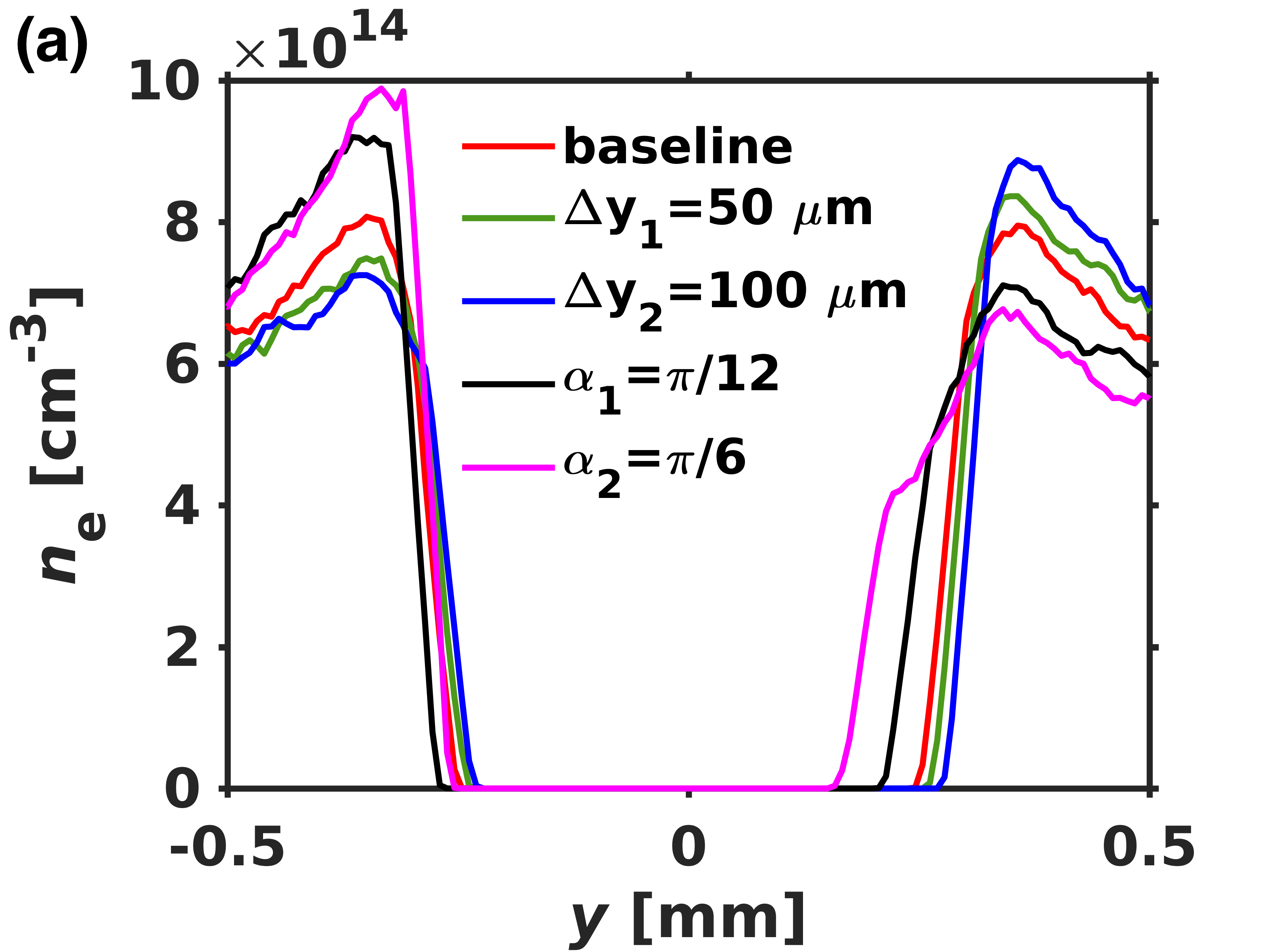}
  }
  \subfloat{
   \centering
   \includegraphics[width=0.47\linewidth]{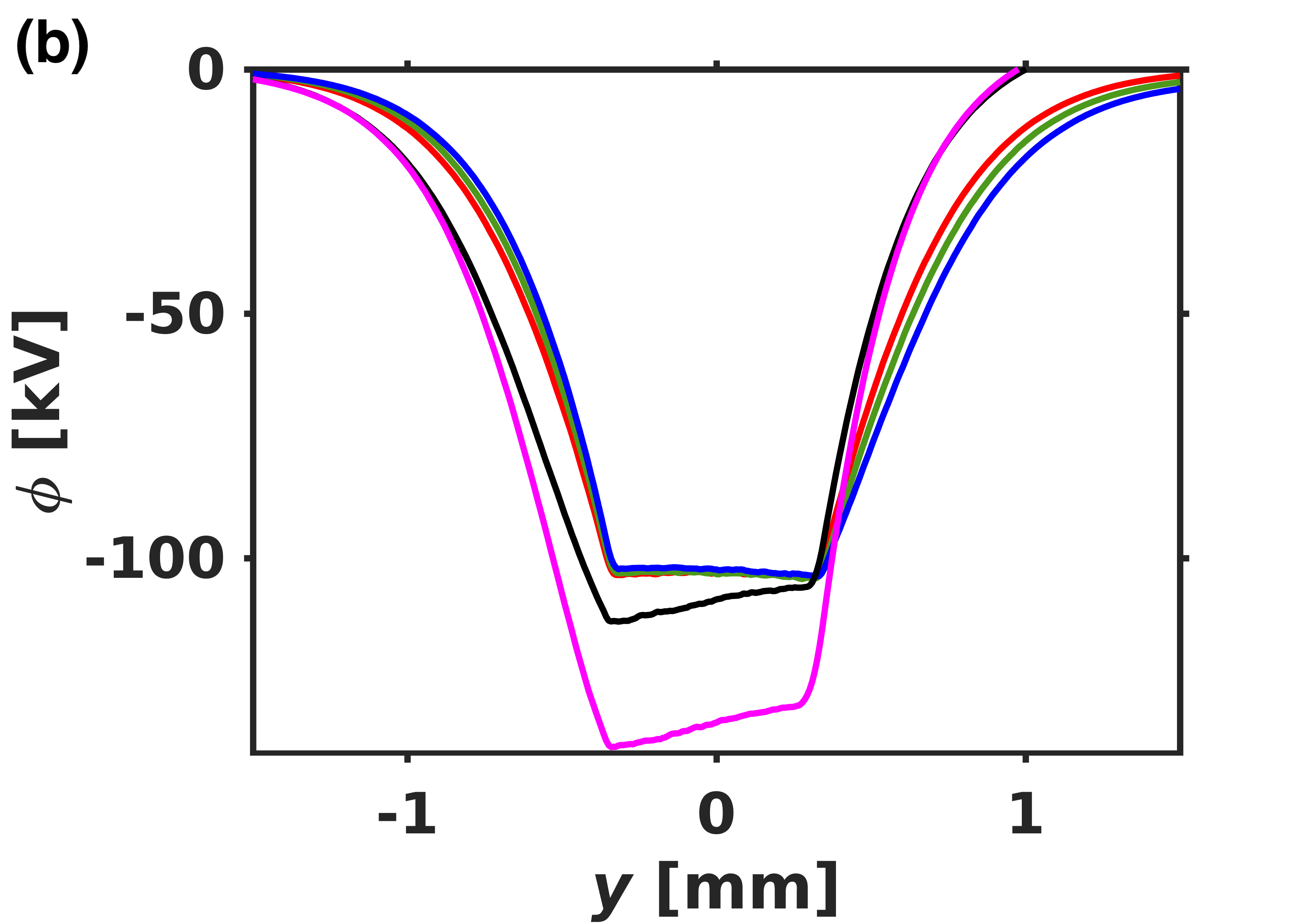}
  }
  \newline
    \subfloat{
   \centering
   \includegraphics[width=0.47\linewidth]{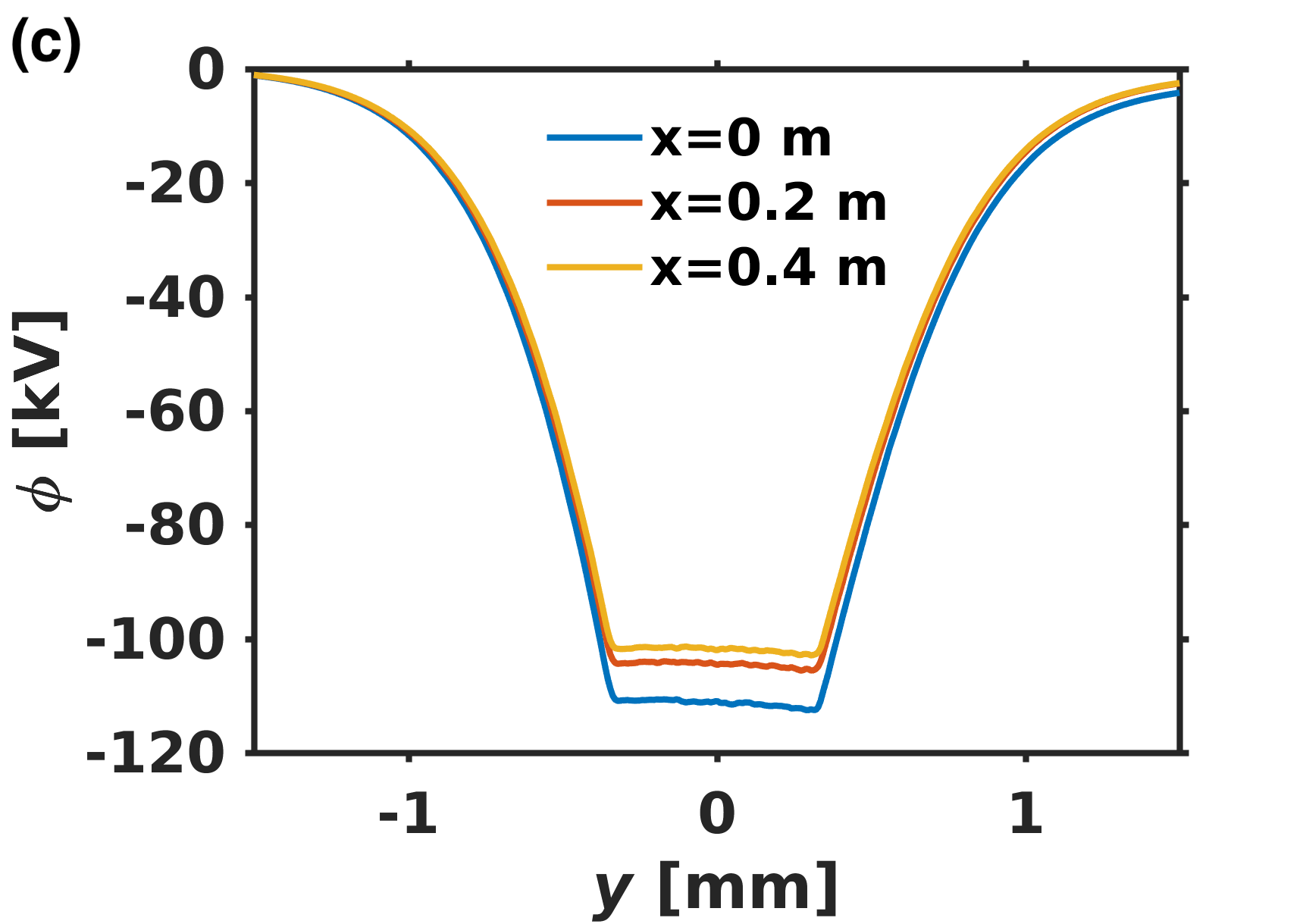}
  }
  \subfloat{
   \centering
   \includegraphics[width=0.47\linewidth]{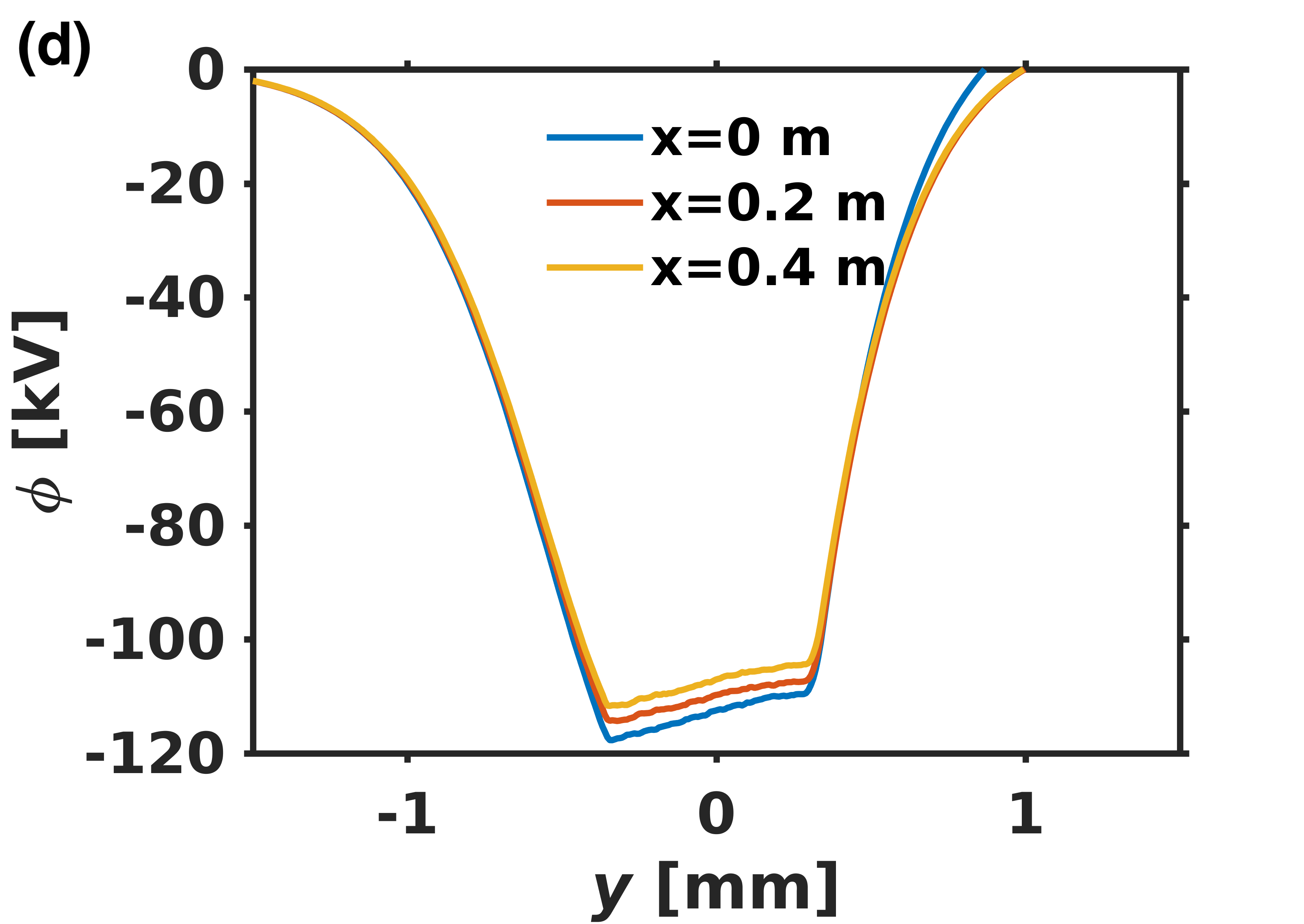}
  }
 \end{center}
 \caption{Plasma electron densities (a) and wake potentials (b) measured along the longitudinal centre of the driver at the distance of 0.4 m for different cases. The same coloured lines in (a) and (b) denote the same case. Here we include two more cases with larger off-axis distance ($\Delta y_2=100~\mu m$) and tilt angle ($\alpha_2=\pi/6$) to enhance the contrast. Wake potentials at different distances for the case $\Delta y_1=50~\mu m$ (c) and $\alpha_1=\pi/12$ (d).}
 \label{DB}
  \vspace*{-1.0\baselineskip}
\end{figure}

Fig.\,\ref{comparison} displays three cases under consideration where the proton driver initially has different states: normal (baseline, Fig.\,\ref{comparison}a), with a positive y offset ($\Delta y$) of 50 $\mu$m (Fig.\,\ref{comparison}d) and with an anti-clockwise and linear tilt angle ($\alpha$) of $pi/12$ (Fig.\,\ref{comparison}g). x and y are the axial and radial coordinates respectively. In hollow plasma, the protons bunch is free from defocusing by ions within the channel and strongly confined by plasma electrons concentrated at the channel boundaries (Fig.\,\ref{comparison}b-c), which form a reflecting-wall like focusing structure (i.e., a deep potential well as shown in Fig.\,\ref{DB}b).

If the proton driver is misaligned from the hollow channel axis, the plasma electron density perturbation at the driver will be asymmetric along the axis. For a positive beam-channel offset of 50 $\mu$m, more plasma electrons concentrate at the upper sheath of the first bubble (Fig.\,\ref{comparison}e), where there is a higher peak density in comparison with the baseline case (Fig.\,\ref{DB}a). The peak density in the other side of the bubble is smaller. Change of the bubble leads to the variation and asymmetry of the transverse plasma wakefields (Fig.\,\ref{comparison}f), which will apparently alter the transverse evolution of the proton bunch. From Figs.\,\ref{DB}b-c, we see the potential bottom is slightly inclined along the beam offset direction, which suggests that the protons within the channel tend to accumulate at the upper side of the bubble and thus the beam centroid gradually moves towards the positive y direction.   

In the tilted driver case, we deliberately pick up an anti-clockwise tilt to create plasma density perturbation in an opposite way (Fig.\,\ref{DB}a). More specifically, the plasma density has a larger peak density but at the down sheath of the bubble (Fig.\,\ref{comparison}h) and the potential bottom is oblique towards the negative y direction (Figs.\,\ref{DB}b and \ref{DB}d). In comparison with the beam offset case, the accelerating gradient decreases considerably with the beam tilt as which equivalently extends the beam length (Fig.\,\ref{WB}a). Also the tilted beam change the wake potential more significantly. 

Based on the above discussion, we can conclude that without the defocusing plasma ions within the hollow channel, the proton driver is capable of sustaining a relatively large offset (100 $\mu$m) or tilt ($\pi$/6) without drastic distortion of the beam shape or beam hosing but the bunch centroid could deviate observably in a long distance. 

\subsection{Wake Characteristics in the Witness Area}

In this section, we focus on the wake characteristics in the second bubble where the witness bunch is accelerated. Note that the witness bunch is not introduced in all simulations, as there is no quadrupole module adopted so the electron bunch if injected will diverge even in the baseline case \cite{item:4, item:5}. 

\begin{figure}[!h]
  \vspace*{-.5\baselineskip}
 \begin{center}
   \subfloat{
   \centering
   \includegraphics[width=0.5\linewidth]{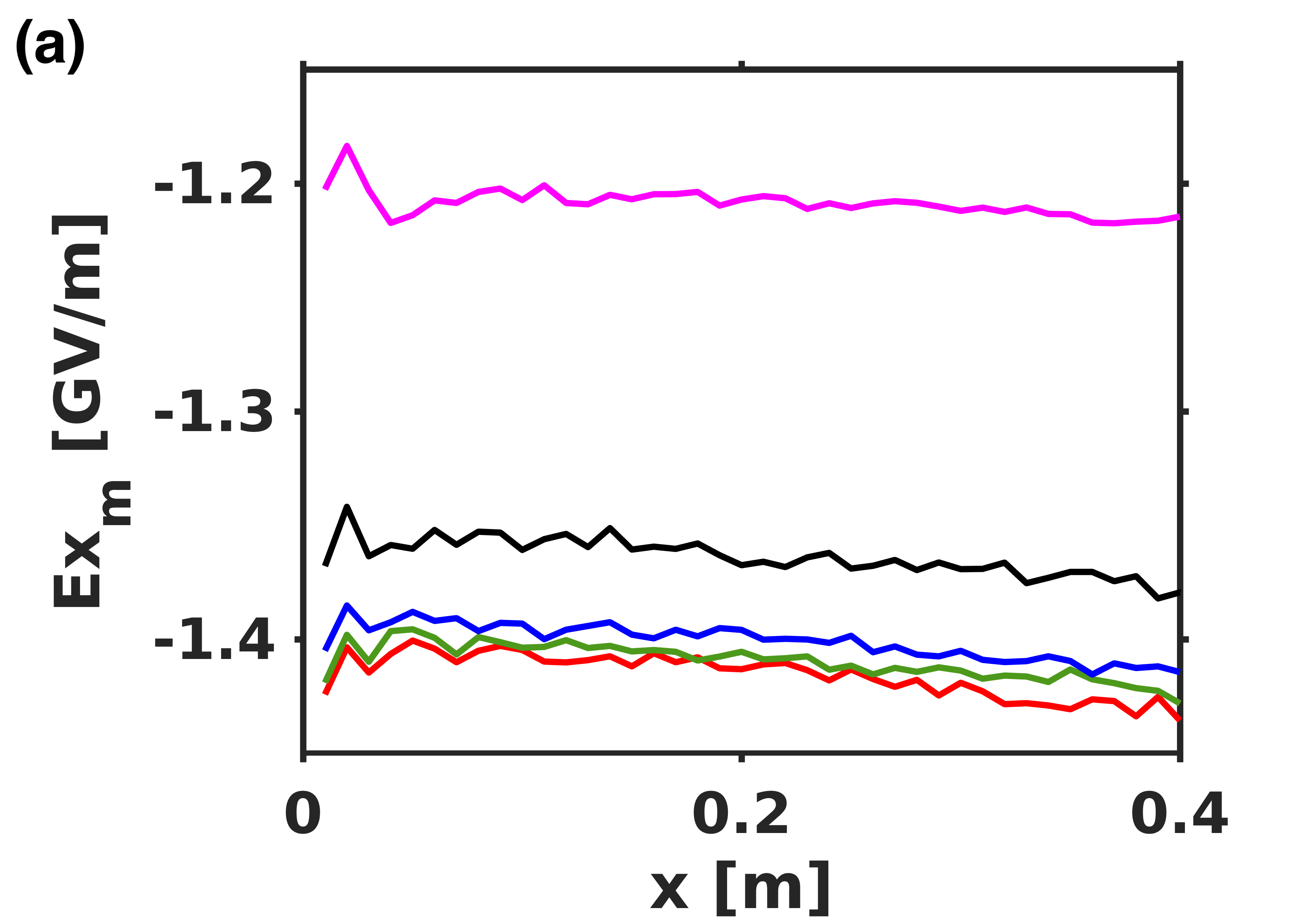}
  }
  \subfloat{
   \centering
   \includegraphics[width=0.5\linewidth]{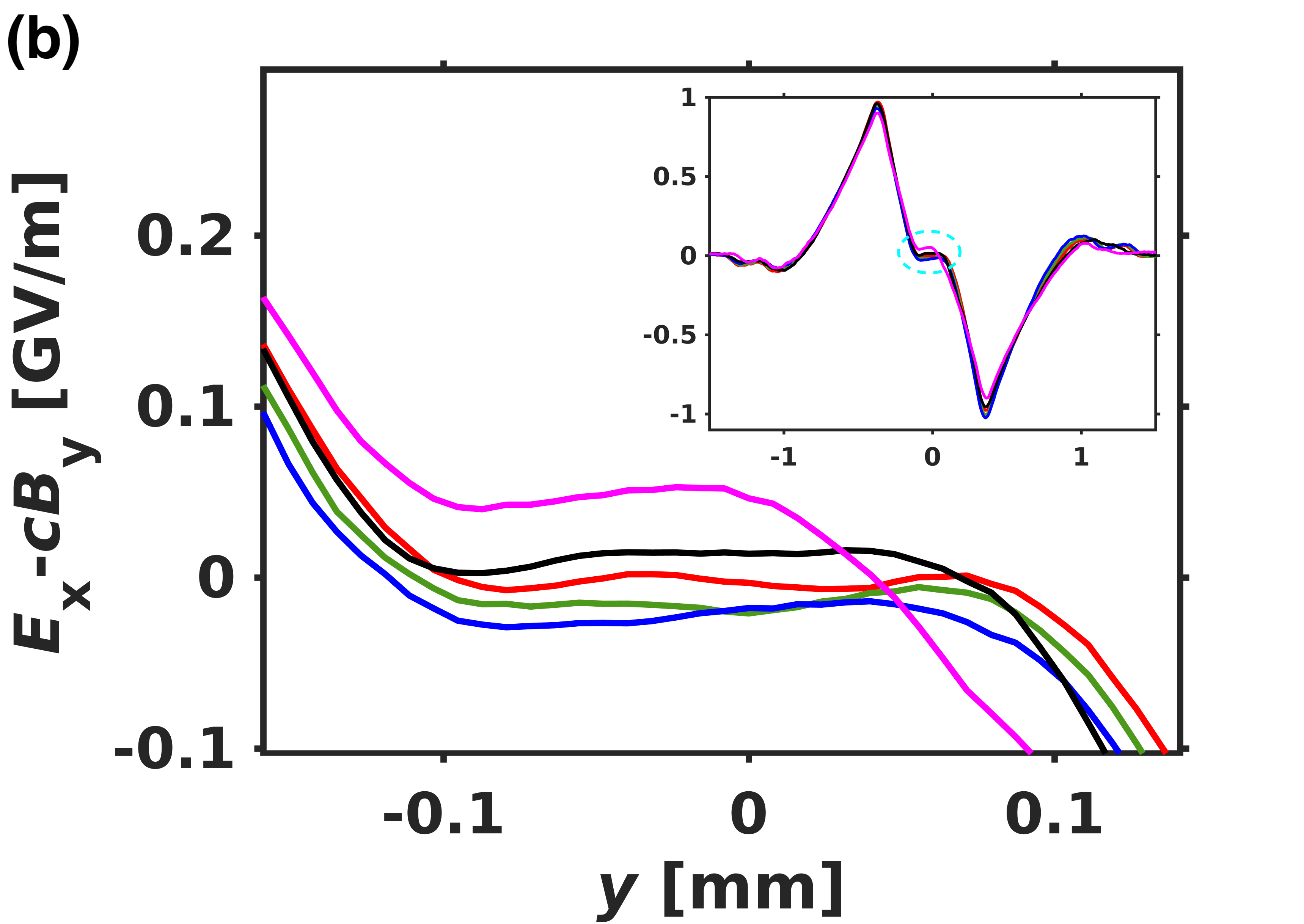}
  }
  \newline
  \subfloat{
   \centering
   \includegraphics[width=1\linewidth]{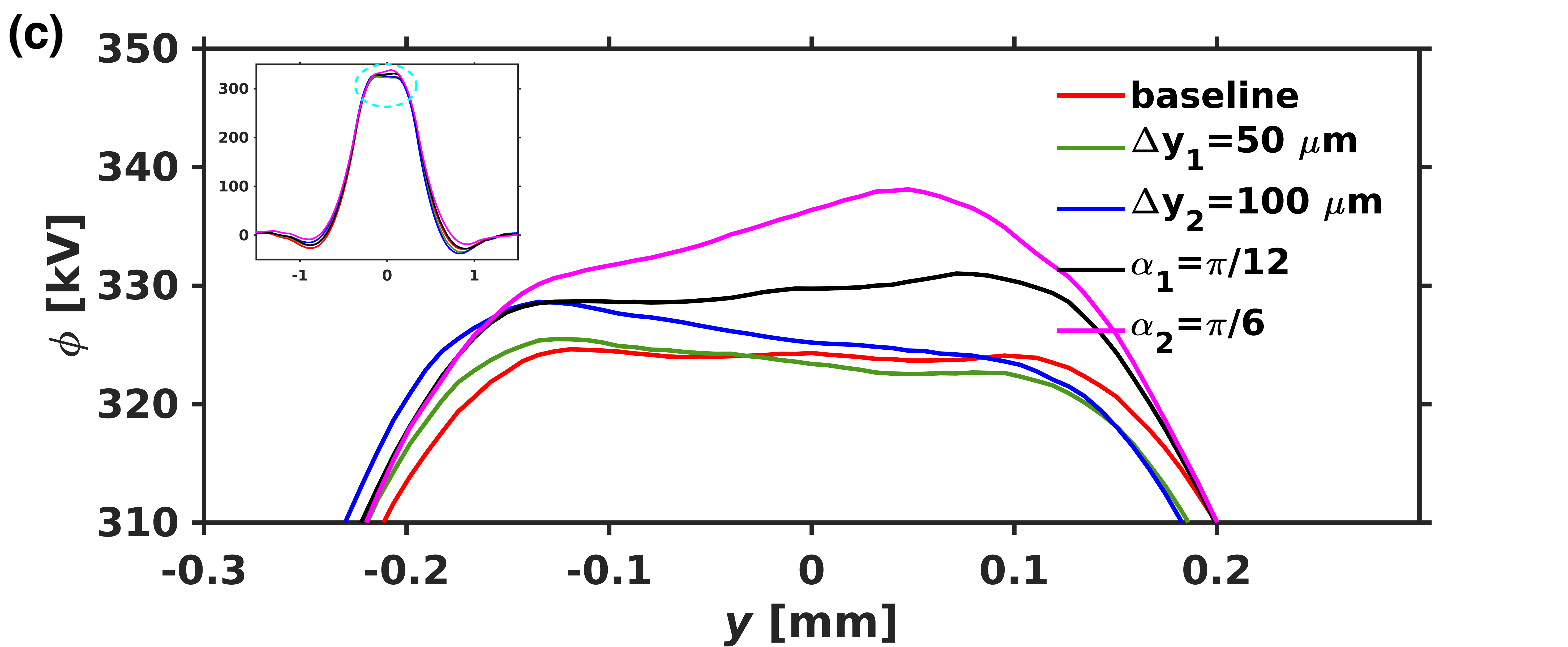}
  }
 \end{center}
  \caption{Maximum accelerating field vs. propagation distance (a), transverse plasma wakefields (b) and wake potentials (c) measured at the longitudinal centre of the witness bunch for different driver cases. The insets show the plots in large ranges. The same coloured lines denote the same case.}
 \label{WB}
  \vspace*{-0.5\baselineskip}
\end{figure}

The second column of Fig.\,\ref{comparison} shows that the offset or tilt of the driving bunch causes less asymmetry of the second bubble in comparison with the first one. Due to uneven plasma electron density distribution, the transverse wakefield becomes negative instead of zero along the radius within the range of a potential witness bunch (Fig.\,\ref{WB}b), which therefore will diffract the witness electrons if located. Take the case of offset $\Delta y_1=50 \mu$m for instance, the transverse plasma wakefield is around 15 MV/m. Fig.\,\ref{WB}c gives the wake potential along the radius at the witness bunch. The witness bunch is located in an area where the potential is much larger than that in large radii. With beam offset or tilt, the witness bunch is located in a region where the potential is radially oblique. As a result, the witness bunch will be pushed by the transverse plasma wakefields towards the positive (negative) y direction until it gets out of the accelerating bubble for the offset (tilt) case. 

\subsection{Mitigation of the Transverse Diffraction}
From the former sections, we can conclude that the proton driver is robust to the misalignment or tilt; nevertheless, the witness bunch is easily deflected by the wake change induced. Here we propose some solutions to confine the witness bunch. The first suggestion is to place the witness bunch closer to the driver so that it resides in the region where there is strong focusing at the hollow boundaries (Figs.\,\ref{comparison}c, 1f, and 1i). As a consequence, the witness bunch can be focused within the channel under initial driver offset or tilt. This is an advantage in comparison with the linear interaction case in the hollow plasma structure \cite{item:7}. The problem is the beam quality might degrade a lot when the witness bunch keeps being reflected by the boundaries. The second way is to add enough quadrupole focusing for the witness bunch. In this section, we assess in detail the third way, which is to adopt a near-hollow plasma structure.        

Near-hollow plasma assumes plasma with a lower density ($n_{in}$) in the channel than in the outer annular plasma ($n_e$), so there will be focusing from plasma ions located within the channel. But a low witness beam emittance is still obtainable as the transverse plasma focusing is weak. In theory, the focusing electric field generated by uniform plasma ions follows the equation of $E_r=n_0er/(2\epsilon_0)$, where $n_0$ is the ion density and $\epsilon_0$ is the vacuum permittivity. Assuming the witness bunch radius is 50 $\mu$m, to get focusing field which can compensate the radial field at the witness bunch caused by the driver offset of 50 $\mu$m, the plasma density is around 3~$\times$~$10^{13}$ cm$^{-3}$, which is almost 0.1$n_e$.   

\begin{figure}[!h]
 \vspace*{-0.5\baselineskip} 
 \begin{center}
   \centering
   \includegraphics[width=0.85\linewidth]{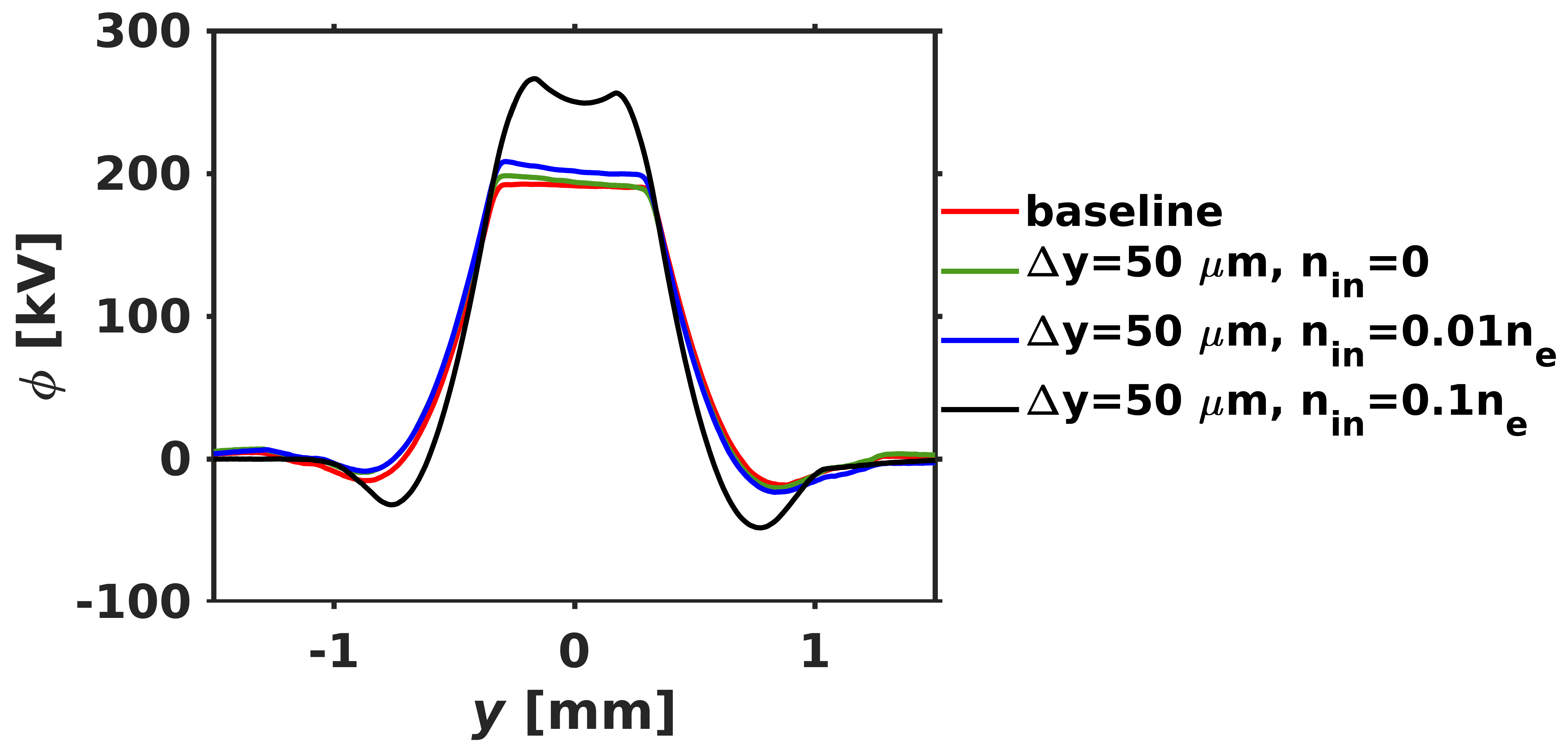}
 \end{center}
 \caption{Wake potentials at the witness bunch for different cases when the driver travels for 0.4 m.}
 \label{near-hollow}
  \vspace*{-1\baselineskip}
\end{figure}

Fig.\,\ref{near-hollow} demonstrates that by employing near-hollow plasma where $n_{in}=0.1n_e$, a potential well at the witness bunch is created. As a result, the witness bunch can be confined under driver misalignment provided that its transverse kinetic energy is smaller than the potential well energy. Here we also give the case where $n_{in}=0.01n_e$ for comparison. We can see that the wake potential is similarly oblique as the beam offset case in hollow plasma. The reason is apart from plasma ions, near-hollow plasma also introduces extra plasma electrons initially located within the channel. These plasma electrons participate in the wake excitation as well, so the perturbed plasma density difference between two sides of the bubble brings extra defocusing wakefields to the witness bunch. 

\section{CONCLUSIONS}
In this paper, we have demonstrated some preliminary results regarding the initially misaligned or tilted driving proton bunch traveling in the hollow plasma. The proton bunch itself is less sensitive to the initial distortion, but the resultant asymmetry of the accelerating bubble drastically lead to the divergence and then loss of the witness bunch. The potential solutions are addition of quadrupoles, change of the witness bunch loading location and adopting a near-hollow plasma, which has the potential to conserve the witness quality. A quasi-static PIC code is essential to simulate the cases discussed for longer distance to further examine the effect of the beam misalignment or tilt.

\iffalse  
	\newpage
	\printbibliography

\else


\null  

\fi

\end{document}